\documentclass[prl,twocolumn,twoside,showpacs,floatfix]{revtex4}
\usepackage{epsf}
\begin{document}
\newcommand{\be}{\begin{equation}}
\newcommand{\ee}{\end{equation}}
\newcommand{\kin}{k_{\rm in}}
\newcommand{\kout}{k_{\rm out}}
\draft
\title{Robustness of the in-degree exponent for the world-wide web\\}
\author{B.~Kahng$^1$, Y. Park$^2$, and H.~Jeong$^3$\\} 
\affiliation{\mbox{$^1$ School of Physics and Center for Theoretical Physics,
Seoul National University, Seoul 151-747, Korea} \\
\mbox{$^2$ Department of Physics, Myongji University, Yongin, Kyunggi-do
449-728, Korea} \\
\mbox{$^3$ Department of Physics, Korea Advanced Institute of Science and 
Technology, Taejon 305-701, Korea}\\ } 
\date{\today}
\begin{abstract}
We consider a stochastic model for directed scale-free networks following 
power-laws in the degree distributions in both incoming 
and outgoing directions. In our model, the number of 
vertices grow geometrically with time with growth rate $p$. 
At each time step, (i) each newly introduced vertex is connected 
to a constant number of already existing vertices with the 
probability linearly proportional to the in-degree of a selected vertex, 
and (ii) each existing vertex updates its outgoing edges 
through a stochastic multiplicative process 
with mean growth rate of outgoing edges $g$ and variance 
$\sigma^2$. 
Using both analytic treatment and numerical simulations, we show 
that while the out-degree exponent $\gamma_{\rm out}$ depends on 
the parameters, the in-degree exponent $\gamma_{\rm in}$ has 
two distinct values, $\gamma_{\rm in}=2$ for $p > g$ and 1 for $p < g$, 
independent of different parameters values.  The latter case 
has logarithmic correction to the power-law. Since the 
vertex growth rate $p$ is larger than the degree growth rate $g$ 
for the world-wide web (www) nowadays, the in-degree exponent 
appears robust as $\gamma_{\rm in}=2$ for the www. 
\end{abstract}
\maketitle
\section{I. Introduction}
Complex system consists of many constituents such as individuals, 
substrates, and companies in social, biological, and economic 
systems, respectively, showing cooperative phenomena between 
constituents through diverse interactions and adaptations to 
the pattern they create\cite{nature,science}. Recently there 
have been considerable efforts to understand such complex systems 
in terms of random graph, consisting of vertices and edges, 
where vertices (edges) represent constituents (their interactions). 
This approach was initiated by Erd\"os and R\'enyi (ER)\cite{er}. 
In the ER model, the number of vertices is fixed, while edges 
connecting one vertex to another occur randomly with certain 
probability. The ER model is however too random to describe 
complex systems in real world.\\  

An interesting feature emerging in such complex systems is 
the scale-free (SF) behavior in the degree distribution, 
$P(k)\sim k^{-\gamma}$, where the degree $k$ is the number 
of edges incident upon a given vertex. 
Barab\'asi and Albert (BA)\cite{physica,ba} introduced an 
evolving model illustrating SF network. In the BA model, 
the number of vertices increases linearly with time, and a 
newly introduced vertex is connected to $m$ already existing 
vertices, following the so-called preferential attachment (PA) 
rule that the vertices with more 
edges are preferentially selected for the connection to the 
new vertex with the probability linearly proportional 
to the degree of that vertex. Then it is known that the 
degree distribution follows $P(k)\sim k^{-3}$ for the BA model.  
While the BA model is meaningful as the first step to generate 
SF network, it is too simple to be in accordance 
with the real-world networks. Extended versions of the BA model 
have been introduced\cite{reka,ginestra}, taking into account 
of additional local events such as adding new edges, or rewiring 
edges from one vertex to another. Depending on the frequency of 
these processes, the degree distribution either remains as SF  
with the exponent depending on the details of the local event or 
follows an exponential decay. \\

Huberman and Adamic (HA)\cite{ha} proposed another scenario for 
SF networks. In the HA model, the number of vertices grows 
geometrically with time, and edges of each vertex evolve 
following a stochastic multiplicative process. Combining these 
two ingredients leads to a power-law behavior in the degree 
distribution, where the exponent is determined by the growth 
rates of vertices, and the mean degree and variance 
of the fluctuations arising in the stochastic process of updating 
edges. While the HA and BA models 
look fundamentally different at a fist glance, they are similar in 
essence. One can show easily that the multiplicative 
process is reduced to the PA rule when the time dependence of 
the total number of edges is the same as that of the number of 
newly introduced vertices. Moreover, the stochastic process in 
the HA model might be related to the rewiring process in the 
extended model of the BA model\cite{reka}.\\ 
 
SF networks may be classified into undirected or directed
network whether the directionality is assigned to edges or not.
Typical examples of undirected networks include the actor 
network\cite{actor}, the author collaboration network\cite{coworker}, 
and the Internet with equal uploading and downloading rates\cite{internet}. 
Directed networks are also ubiquitous in real world such as 
the world-wide web (www)\cite{ha,albert,kumar}, the citation 
network of scientific papers\cite{redner}, biological networks 
such as metabolic networks\cite{metabolic} and neural networks, $etc$. 
Recently, Albert $et$ $al.$\cite{albert} and Huberman $et$ $al.$
\cite{ha} investigated the topology of the www extensively, and 
found that the in-degree and the out-degree distributions of the 
www exhibit power-law behaviors with different exponents, i.e., 
$P_{\rm in}(\kin)\sim \kin^{-\gamma_{\rm in}}$ and 
$P_{\rm out}(\kout)\sim \kout^{-\gamma_{\rm out}}$, respectively.  
Here the in-degree $k_{\rm in}$ (out-degree $k_{\rm out}$) means 
the number of edges incident upon (emanating from) a given vertex. 
Further studies\cite{kumar,Kleinberg,Broder} showed that 
$\gamma_{\rm in}$ is robust as $\gamma_{\rm in}\approx 2.1$ 
for different systems, while $\gamma_{\rm out}$ varies depending 
on systems in the range, $2.4\sim 2.7$.\\ 

Theoretical studies for directed networks have less been 
carried out compared with those for undirected networks. 
When the directionality is assigned to edge in the BA model, pointing 
from a new vertex to old ones, the in-degree and the out-degree 
distributions follow $P_{\rm in}(\kin)\sim \kin^{-3}$ and 
$P_{\rm out}(\kout)=\delta(\kout-m)$ respectively, which 
is not relevant to the empirical results for the www. Dorogovtsev 
and Mendes\cite{Mendes} performed a similar study using the rate 
equation, in which the in-degree distribution follows a power-law 
whereas the out-degree distribution is of the $\delta$-function.
More recently, Krapivsky $et$ $al.$ \cite{krapivsky} studied 
directed SF networks using the rate equation method 
for the simple case similar to the one introduced by Tadi\'c\cite{tadic}
that at each time step, a vertex is newly introduced and connected 
to an old vertex following the PA rule with a certain probability 
and an internal directed edge is connected between two vertices 
chosen following the PA rule with the remaining probability.  
They obtained the in-degree and the out-degree distributions 
analytically, both of which exhibit power-law behaviors with different 
exponents depending on the detail of the parameters they used. 
While their analytic treatment was successful in generating the 
empirical values of the out-degree and the in-degree exponents 
for the www by tuning the parameters, their model is unable to 
illustrate the robustness of the in-degree exponent for various 
systems because tuning parameters leads to different values of  
of $\gamma_{\rm in}$ and $\gamma_{\rm out}$ at the same time.\\

In this paper, we introduce a stochastic model for directed SF networks 
exhibiting power-law behaviors with distinct exponents 
in both incoming and outgoing directions and present an analytic 
solution for the model. Through this study, we can illustrate 
why the in-degree exponent is robust for different 
systems, while the out-degree exponent depends on the details 
of systems. This behavior occurs when the growth rate of the number 
of vertices is large enough compared with the effective growth rate 
of degree of each vertex.\\ 
  
This paper is organized as follows. In section II, we will introduce 
a stochastic model. In sections III and IV, analytic solutions 
for the out-degree and the in-degree distributions will be presented, 
respectively. In section V, we will present the result of numerical 
simulations for the model in the vertex growth dominant and the degree 
growth dominant regimes, respectively. The final section will be 
devoted to the conclusions. 

\section{II. The model}

Let us introduce a directed SF network model as follows: 
(i) At each time step, the total number of vertices increases 
geometrically with growth rate $p$, i.e., 
\begin{equation}
N(t)=N(t-1)(1+p).
\end{equation}
So the total number of vertices newly introduced at time $t$ is 
$pN(t-1)$. (ii) $m$ edges emanate from each new vertex, 
pointing to $m$ distinct old vertices following 
the PA rule. The probability to connect to a vertex $j$ is given by  
\begin{equation}
\Pi_{i\rightarrow j}={{k_{{\rm in}, j}(t-1)} \over 
{\sum_{r=1}^{N(t-1)} k_{{\rm in},r}(t-1)}}, 
\label{pa}
\end{equation}
where $k_{{\rm in},j}(t-1)$ means the in-degree of the vertex
$j$ at time $t-1$. 
We assume in the model that each new vertex is given an incoming 
edge pointed from itself, otherwise in-degree never grows with time. 
(iii) each vertex updates its outgoing edges by either 
adding new edges or deleting existing edges through a multiplicative 
stochastic process. Let $k_{{\rm out},i}(t)$ denote the out-degree 
of vertex $i$ at time $t$. Then $k_{{\rm out},i}(t)$ evolves as
\begin{equation}
k_{{\rm out},i}(t+1)=k_{{\rm out},i}(t)(1+\zeta_i(t+1)),
\label{multi}
\end{equation}
where $\zeta_i(t)$ means the growth rate of the out-degree
$k_{{\rm out},i}(t)$ at time $t$, which fluctuates from time 
to time about mean $g_i$, 
\begin{equation}
\zeta_i(t)=g_i+\xi_i(t),
\end{equation}
where $\xi_i(t)$ is assumed to be a white noise satisfying 
$\langle \xi_i(t) \rangle=0$ and 
$\langle \xi_i(t)\xi_j(t') \rangle =\sigma_i^2 \delta_{t,t'}
\delta_{i,j}$, where $\sigma_i^2$ is the variance. 
The growth rate $g_i$ and the standard deviation $\sigma_i$ 
could vary in general for different vertices. HA, however, assumed 
that $\{\zeta_i\}$ are uniform for different vertices, i.e., 
$g_i=g$ and $\sigma_i=\sigma$ for all $i$. 
When $\zeta_i(t+1) > 0$, the out-degree at 
vertex $i$ is increased. Then we add $k_{{\rm out},i}(t)\zeta_i(t+1)$ 
new edges to the vertex $i$, pointing to other distinct 
vertices which are not connected, according to the PA rule given by 
Eq.(\ref{pa}). When $\zeta_i(t+1)<0$, we delete 
$k_{{\rm out},i}(t)|\zeta_i(t+1)|$ outgoing edges 
from the vertex $i$ randomly.

\section{III. The out-degree distribution}

The out-degree distribution $P_{\rm out}(\kout)$ can be obtained 
by following the argument given by HA. 
The conditional probability $P_{\rm out}(\kout,\tau~|~m)$ 
that $k_{{\rm out},i}=\kout$ at time $t=t_i+\tau$ for a vertex $i$ 
born at $t=t_i$ with $k_{{\rm out},i}=m$ is given by 
\begin{eqnarray}\label{gaussian}
P_{\rm out}&&\Big(\kout, \tau | m \Big)= \nonumber \\
&&{1\over {\kout\sqrt{2\pi \sigma_0^2 \tau}}} 
\exp\Big\{-{(\ln \big(\kout/m\big)- g_0\tau)^2 
\over {2\sigma_0^2\tau}}\Big\}.
\end{eqnarray}
The above distribution was obtained by applying the central limit 
theorem for the variable $\ln (k_{\rm out}(t)/k_{\rm out}(t-1))$, 
so that $g_0$ and $\sigma_0^2$ in Eq.(\ref{gaussian}) are related to 
$g$ and $\sigma^2$ as $g_0 \approx g-\sigma^2/2$, and 
$\sigma_0^2 \approx \sigma^2$, respectively\cite{gauss}. 
Since the density of vertices with age $\tau$ is proportional to
$\rho(\tau)\sim \exp(-p\tau)$, the out-degree distribution collected 
over all ages becomes
\begin{equation}
P_{\rm out}(\kout)=\int d\tau \rho(\tau)P_{\rm out}(k_{\rm out},\tau | 
m) \sim \kout^{-\gamma_{out}},
\end{equation}
where
\begin{equation}
\gamma_{\rm out}=1-\frac{g_0}{\sigma_0^2}+
\frac{\sqrt{g_0^2+2p\sigma_0^2}}{\sigma_0^2}.
\label{outdegree}
\end{equation}
We note that the out-degree exponent $\gamma_{\rm out}$ depends on 
the three parameters, $p$, $g_0$ and $\sigma_0$.

\section{IV. The in-degree distribution}

The in-degree at a vertex $i$ is increased as new edges are 
additionally pointed from other vertices to $i$, or decreased as already 
connected edges are deleted from other vertices. 
For the increased case, there are two types of occasions. 
The first is the case that some of edges from newly born 
vertices are connected to the vertex $i$. 
Since the total number of edges generated from new vertices 
at time $t$ is given by 
\begin{equation}
{\cal{L}}_{\rm new}(t)=mpN(t-1),
\end{equation}
the in-degree of the vertex $i$ evolves as  
\begin{equation}
\frac{\partial k_{{\rm in},i}(t)}{\partial t}=
{{k_{{\rm in},i}(t-1)}\over{\sum_{r=1}^{N(t-1)}k_{{\rm in},r}(t-1)}}
{\cal {L}}_{\rm new}(t). 
\label{new}
\end{equation}
Second is the case that the vertex $i$ receives edges from existing 
vertices as they update their outgoing edges. The total number 
of newly added outgoing edges is given by 
\begin{equation}
{\cal{L}}_{\rm add}(t)=\sum_{j=1}^{N(t-1)}k_{{\rm out},j}(t-1)
\zeta_j^{+}(t),
\label{add}
\end{equation}
where $\zeta_j^+(t)$ denotes the one when $\zeta_j(t)>0$. 
Then, the in-degree of the vertex $i$ evolves as  
\begin{equation}
\frac{\partial k_{{\rm in},i}(t)}{\partial t}=
{{k_{{\rm in},i}(t-1)}\over{\sum_{r=1}^{N(t-1)}k_{{\rm in},r}(t-1)}}
{\cal {L}}_{\rm add}(t). 
\label{add}
\end{equation}
On the other hand, the decreased case occurs when other vertices 
remove their connections to the vertex $i$. This case occurs when 
$\zeta_j (t) <0$ for a vertex $j\ne i$, with $\zeta_j(t)$ denoted
by $\zeta_j^{-}(t)$. The total number of edges removed through this 
updating process is  
\begin{equation}
{\cal{L}}_{\rm del}(t)=\sum_{j=1}^{N(t-1)}k_{{\rm out},j}(t-1)
|{\zeta}_j^{-}(t)|.
\label{del}
\end{equation}
Although the edges deleted are chosen randomly,  
the vertex with more in-degree has more incoming edges deleted 
because incoming edges were formed following the PA rule. 
Thus the deletion process leads to  
\begin{equation}
\frac{\partial k_{{\rm in},i}(t)}{\partial t}=
{{k_{{\rm in},i}(t-1)}\over{\sum_{r=1}^{N(t-1)}k_{{\rm in},r}(t-1)}}
{\cal {L}}_{\rm del}(t). 
\label{del}
\end{equation}
Altogether the dynamic equation for the in-degree of the vertex $i$ is 
written as 
\begin{eqnarray}
\frac{\partial k_{{\rm in},i}(t)}{\partial t} &=&
{{k_{{\rm in},i}(t-1)}\over{\sum_{r=1}^{N(t-1)}k_{{\rm in},r}(t-1)}}
\Big({\cal {L}}_{\rm new}(t)\nonumber \\ \nonumber \\ &&
+{\cal {L}}_{\rm add}(t)-{\cal {L}}_{\rm del}(t)\Big). \label{deltak}
\end{eqnarray}
The above equation can be rewritten as 
\begin{equation}
\frac{\partial k_{{\rm in},i}(t)}{\partial t}
=k_{{\rm in},i}(t-1)\Big(\frac{m p N(0)e^{p t } }{L(t)}
+g_0+\frac{\chi(t)}{L(t)} \Big),
\label{dynamic}
\end{equation}
where $L(t)$ denotes the total number of incoming edges at time $t$,   
\begin{equation}
L(t)=\sum_i^{N(t)}k_{{\rm in},i}(t), 
\end{equation}
which behaves asymptotically as 
\begin{equation}
L(t) \approx  \cases{ A_1 e^{pt}, & if $p > g_0$, \cr
              A_2 t e^{pt}, & if $p=g_0$, \cr
              A_3 e^{g_0 t}, & if $p < g_0$, }
\end{equation}
where $A_1$, $A_2$, and $A_3$ are given as  
\begin{eqnarray}
A_1 &=& \frac{mp N(0)}{(p-g_0)},\\ A_2 &=& mp N(0) ,\\ 
\noalign{\hbox{and}} A_3 &=& \frac{mpN(0)}{(g_0-p)}.
\end{eqnarray}
$\chi(t)$ in Eq.(\ref{dynamic}) is defined as  
\begin{equation}
\chi(t)=\sum_i^{N(t-1)} k_{{\rm out},i}(t-1) \Big(\xi_i^+-|\xi_i^-|\Big),
\end{equation}
where $\xi_i^+(t)$ ($\xi_i^-(t)$) denotes the noise for 
$\xi_i(t) >0$ ($\xi_i(t) <0$). 
Then using the stochastic property, $\langle \xi_i \rangle=0$, 
we obtain that
\begin{equation}
\langle \chi(t) \rangle =0,
\end{equation}
and
\begin{equation}
\langle \chi(t)\chi(t')\rangle \approx \cases{ B_1
e^{pt}\delta_{t,t'}, & if $p > 2(g_0+\sigma_0^2/2)$, \cr B_2 t
e^{pt}\delta_{t,t'}, & if $p=2(g_0+\sigma_0^2/2)$, \cr B_3
e^{2(g_0+\sigma_0^2/2)t}\delta_{t,t'}, & if $p < 2(g_0+\sigma_0^2/2)$,}
\end{equation}
where $B_1$, $B_2$, and$B_3$ are given by 
\begin{eqnarray}
B_1 &=& \frac{m^2\sigma_0^2 p N_0}{2(p-\sigma_0^2-2g_0)}, \\
B_2 &=& m^2\sigma_0^2 p N_0}{(1+p)^2,\\
\noalign{\hbox{and}}
B_3 &=& \frac{m^2\sigma_0^2 p N_0}{2(\sigma_0^2+2g_0-p)}.
\end{eqnarray}
Thus $\chi(t)$ plays a role of noise, and its variance depends on 
time.\\
 
The asymptotic behavior of the dynamic equation Eq.(\ref{dynamic}) 
depends on relative magnitudes among $p$, $g_0$, 
and $\sigma_0^2$. We consider every possible case below.\\
 
(I) When $p \ge g_0+{\sigma_0}^2/2$ (i.e., $p \ge g$), the stochastic 
term, the last term in Eq.(\ref{dynamic}), is negligible in long time 
limit. Moreover, since $N(t)$ and $L(t)$ have the same time-dependence,
Eq.(\ref{dynamic}) is simply reduced to 
\begin{equation}
\frac{\partial k_{{\rm in},i}(t)}{\partial t}=
p k_{{\rm in},i}(t).
\end{equation}
Thus the in-degree of a vertex $i$ born at time $t=t_i$
becomes
\begin{equation}
k_{{\rm in},i}(t)=e^{p(t-t_i)}.
\end{equation}
Then the in-degree distribution becomes
\begin{eqnarray}
P_{\rm in}(\kin)&=&{{\partial}\over{\partial k_{{\rm in},i}}}
(1-P(\kin > k_{{\rm in},i}))
\left |_{k_{{\rm in},i}=\kin}\right. \\ & =& 
{{\partial}\over{\partial k_{{\rm in},i}}}
\big(-p\frac{m}{k_{{\rm in},i}} \big)\left |_{k_{{\rm in},i}=\kin}\right. 
\propto \kin^{-{\gamma}_{\rm in}}
\end{eqnarray}
with ${\gamma}_{\rm in}=2$.\\

(II) When $g_0 \le p < g_0+\sigma_0^2/2$ (i.e., $g-\sigma^2/2 \le p < g$), 
the dynamic equation is reduced to asymptotically  
\begin{equation}
\frac{\partial k_{{\rm in},i}(t)}{\partial t}
=k_{{\rm in},i}(t)\Big(p+\frac{\chi(t)}{L(t)}\Big).
\label{dynamic2}
\end{equation}
Since ${\langle \chi(t)\rangle}=0$, one may regard the above 
equation as a stochastic log-normal dynamic equation with the variance, 
\begin{equation}
\frac{\langle \chi(t)\chi(t^{\prime})\rangle}{L(t)^2} 
={\cal {D}}_1^2 e^{2s t}\delta_{t, t^{\prime}},
\end{equation}
with ${\cal {D}}_1^2=B_3/{A_1^2}$ and $s=g_0+\sigma_0^2/2-p$.
Since $s>0$, the fluctuation term cannot be ignored.
Invoking the central limit theorem, the conditional probability 
$P_{\rm in}(\kin,\tau |k_{{\rm in},0} )$, that $k_{{\rm in},i}=\kin$ 
at time $t=t_i+\tau$, given $k_{{\rm in},i}=k_{{\rm in},0}=1$ 
at $t=t_i$ becomes,   
\begin{eqnarray}\label{gauss_in}
&&P_{\rm in}\Big(\kin, \tau | k_{{\rm in},0}\Big)=\nonumber \\
&&{1\over {\kin\sqrt{2\pi {\cal {D}}_1^2 (e^{2s\tau}-1)/2s}}} 
\exp \Big\{-{{(\ln \big(\kin/k_{{\rm in},0}\big)-p\tau)^2} 
\over {2{\cal {D}}_1^2 (e^{2s\tau}-1)/2s}}\Big\}.
\end{eqnarray}
So the in-degree distribution can be obtained through
\begin{equation}
P_{\rm in}(\kin)=\int_0^{t}\rho(\tau)P_{\rm in}(\kin, \tau|k_{{\rm in},0}) 
d\tau.
\end{equation}
When $2st \gg 1 $, it can be shown using the saddle point 
approximation that the in-degree
distribution is of the form,
\begin{equation}
P_{\rm in}(\kin)\sim {1\over{\kin(\ln \kin)^{(p/s+1)}}},
\label{g1}
\end{equation}
which is valid as long as $\ln \kin \ll e^{2st}$. When $p=g_0$, 
${\cal {D}}_1^2$ is replaced by $B_3/A_2^2$.\\

(III) For $p<g_0$ (i.e., $p < g-\sigma^2/2$), the dynamic equation 
of $k_{{\rm in},i}(t)$ can be written as
\begin{equation}
\frac{\partial k_{{\rm in},i}(t)}{\partial t}
=k_{{\rm in},i}(t)\Big(g_0+\frac{\chi(t)}{L(t)}\Big).
\label{dynamic2}
\end{equation}
The variance can be written as 
\begin{equation}
\frac{\langle \chi(t)\chi(t^{\prime})\rangle}{L(t)^2} 
={\cal {D}}_2^2 e^{\sigma_0^2 t}\delta_{t, t^{\prime}},
\end{equation}
with ${\cal {D}}_2^2 \equiv B_3/{A_3^2}$. 
Following the same step as used in the second case, we obtain that  
\begin{equation}
P_{\rm in}(\kin)\sim {1\over{\kin(\ln \kin)^{(2g_0/\sigma_0^2+1)}}}.
\label{g2}
\end{equation}\\

Conclusively, when the growth rate of vertex $p$ is larger 
than the effective growth rate of edge, $g_0+\sigma_0^2/2$, 
the in-degree distribution is independent of the detail of 
evolving networks, so that the in-degree exponent 
is robust for different systems, while the out-degree exponent 
depends on the detail. This is the case we observe in the real www
because the number of webpages increases rapidly nowadays, 
whereas average number of hyperlinks does rather at a 
slower rate due to limited space on webpage.    
When the number of webpages is saturated in the future, 
the growth rate $p$ will become moderated with the number of 
hyperlinks $g_0$ much dominant. Then the in-degree distribution 
exhibits a phase transition to the form Eq.(\ref{g1}) or (\ref{g2}), 
implying that the hyperlink is much centralized to a few famous webpages. 
The phase diagram is depicted in Fig.1. 

\section{V. Numerical simulations}

It was reported\cite{Broder} that the Web consisted of $N=203\times 10^6$ 
documents from the point of view of Altavista, and the average 
in-degree and out-degree are ${\bar k}_{\rm in}={\bar k}_{\rm out}=7.22$ 
as of May of 1999, and $N=271\times 10^6$ and 
${\bar k}_{\rm in}={\bar k}_{\rm out}=7.85$ 
as of October of 1999. Based on this data, we estimate very roughly 
the vertex and mean degree growth rates to be 
$p\approx 0.059$ and $g\approx 0.017$ per month, respectively. 
However, the fluctuation strength $\sigma$ is not known. 
Using the estimated values of $p$ and $g$, we perform numerical 
simulations for the stochastic model following the HA idea, 
where the variance $\sigma_0^2$ are chosen to be in the regions 
(I) and (II). The simulation results are compared with the 
theoretical predictions. First, we choose the variances $\sigma_0^2=0.052$ 
and 0.021 belonging to the region (I). As seen in Figs.2, the in-degree 
exponents $\gamma_{\rm in}$ are robust to be $\gamma_{\rm in}\approx 2$ 
for both cases, while the out-degree exponents $\gamma_{\rm out}$ 
are different from each other as $\gamma_{\rm out}\approx 2.7$ 
for $\sigma_0^2 \approx 0.052$ and $\gamma_{\rm out}\approx 3.0$ for 
$\sigma_0^2 \approx 0.146$. The simulation results are close to  
the theoretical predictions according to Eq.(\ref{outdegree}), 
$\gamma_{\rm out}\approx 2.7$ and $\gamma_{\rm out}\approx 3.1$, respectively.
Second, we choose $p=0.010$, $g=0.017$, and $\sigma_0^2=0.041$ 
belonging to the region (II). The power-law behavior for the in-degree 
distribution appears for large $k_{\rm in}$ with the exponent 
$\gamma_{\rm in}\approx 1$ as shown in Fig.3, in agreement 
with the theoretical prediction without the logarithmic 
correction. For the out-degree distribution, the power-law behavior 
is also obtained with the exponent $\gamma_{\rm out}\approx 1.8$, which 
is in agreement with the theoretical value $1.8$ according 
to Eq.(\ref{outdegree}).

\section{VI. Conclusions}
We have introduced a stochastic model for directed SF networks, which evolves 
with time. In our model, the evolution of outgoing edges follows 
the stochastic multiplicative process, while that of 
incoming edges does the preferential attachment.  
With this model, we could illustrate why the in-degree 
exponent for the www is robust, independent of different systems, 
while the out-degree exponent depends on different systems. 
We presented analytic results for both the in-degree and the out-degree 
distribution and confirmed our theoretical predictions by performing 
numerical simulations with the parameter values estimated from the www 
in real world.

\section{Acknowledgements} 
We would like to thank A.-L. Barab\'asi for helpful discussions.
This work was supported by the Korean Research Foundation 
(Grant No. 01-041-D00061).\\

\begin{figure}[b]
\centerline{\epsfxsize=8cm \epsfbox{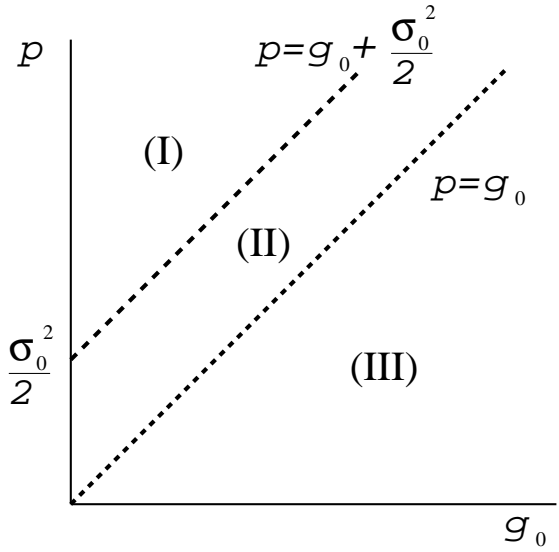}}
\caption{Phase diagram for different behaviors of the in-degree distribution. 
In the region (I), the in-degree distribution shows 
$P(k_{\rm in})\sim k_{\rm in}^{-2}$, while $P(k_{\rm in})\sim 
1/k_{\rm in}(\ln k_{\rm in})^{\beta}$, with 
$\beta=1+p/(g_0+\sigma_0^2/2-p)$ in the region (II), and 
$\beta=1+g_0/\sigma_0^2$ in the region (III).} 
\label{fig:mul}
\end{figure}
\begin{figure}
\centerline{\epsfxsize=8cm \epsfbox{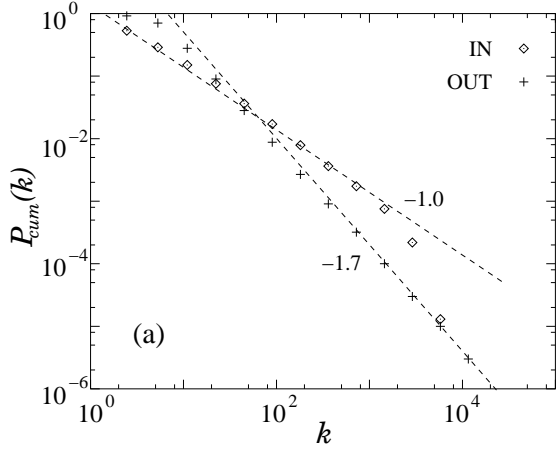}}
\centerline{\epsfxsize=8cm \epsfbox{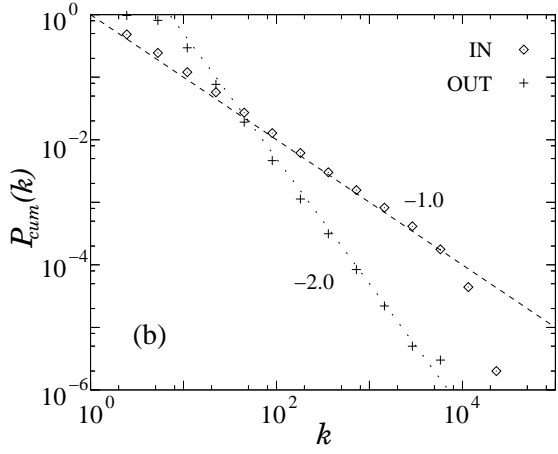}}
\caption{Plot of the in-degree and out-degree distributions drawn in 
the cumulated way, $P_{\rm cum}(k)=\int_k^{\infty} P(k)dk$ for the cases, 
$p=0.059$, $g=0.017$, and $\sigma^2=0.051$ for (a), and $\sigma^2=0.021$ 
for (b). Both cases belong to the region (I).} 
\end{figure}
\begin{figure}
\centerline{\epsfxsize=8cm \epsfbox{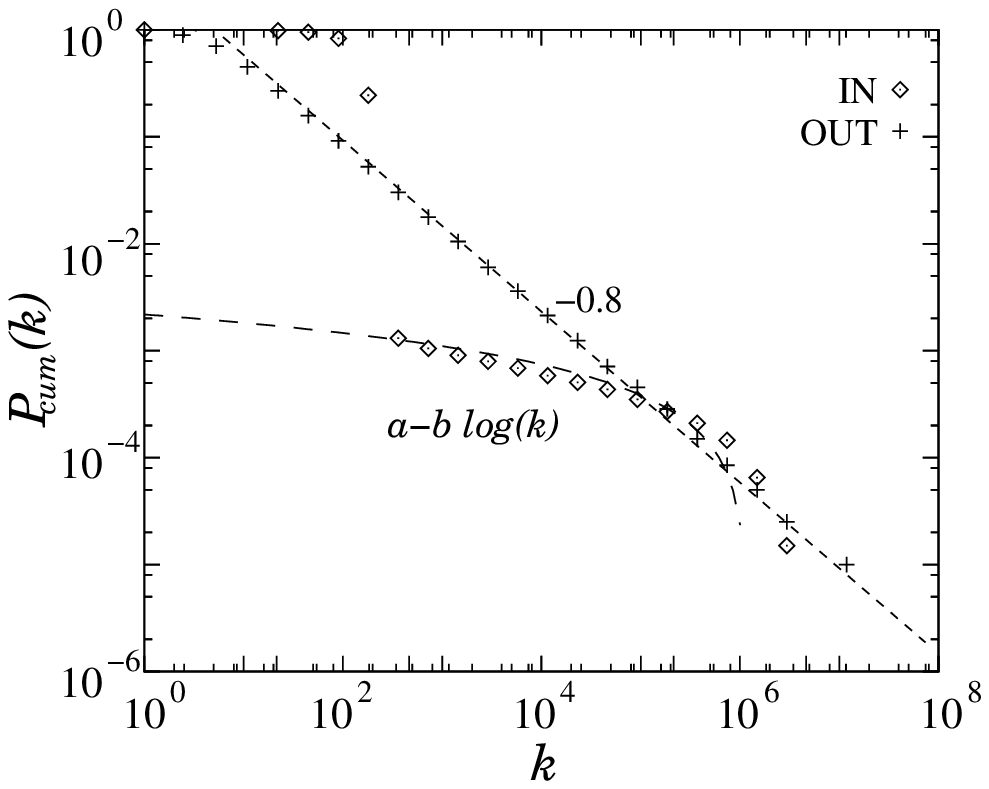}}
\caption{Plot of the in-degree and out-degree distributions drawn in 
the cumulated way, $P_{\rm cum}(k)=\int_k^{\infty} P(k)dk$ for the cases, 
$p=0.010$, $g=0.017$, and $\sigma^2=0.051$, belonging to the region (II). 
The numerical data for the out-degree distribution shows the power-law 
behavior with the exponent $\gamma_{\rm out}\approx 1.8$, and those for 
the in-degree does with $\gamma_{\rm in}\approx 1$.}
\end{figure}


\begin{thebibliography}{99}
\bibitem{nature} S.H. Strogatz, Nature {\bf 410}, 268 (2001). 
\bibitem{science} N. Goldenfeld and L.P. Kadanoff, Science {\bf 284}, 
87 (1999).
\bibitem{er} P. Erd\"os, A. R\'enyl, Publ. Math. Inst. Hung. Acad. Sci. 
{\bf 5}, 17 (1960).
\bibitem{physica}A.-L. Barab\'asi, R. Albert and H. Jeong, 
Physica A,{\bf 272,} 173 (1999); R. Albert and A.-L. Barab\'asi, 
(cond-mat/0106144).
\bibitem{ba}A.-L. Barab\'asi and R. Albert, Science {\bf 286,} 509 (1999).
\bibitem{reka} R. Albert and A.-L. Barabasi, Phys. Rev. Lett. {\bf 85,} 
5234 (2000).  
\bibitem{ginestra} G. Bianconi and A.-L. Barab\'asi, Europhys. Lett.
{\bf 54}, 436 (2001).
\bibitem{ha} B.A. Huberman and L.A. Adamic, Nature {\bf 401,} 131 (1999);
B. A. Huberman and L. A. Adamic, (cond-mat/9901071).
\bibitem{actor} M.E.J. Newman, S.H. Strogatz, and D.J. Watts, 
Phys.Rev.E 64, 026118 (2001).
\bibitem{coworker} M.E.J. Newman, Proc. Natl. Acad. Sci. (USA) {\bf 98,} 
404 (2001); L. A. Amaral, A. Scala, M. Barth\'el\'emy, and H.E. Stanley, 
Proc. Natl Acad. Sci. USA {\bf 97}, 11149 (2000).
\bibitem{internet} E.W. Zegura, K.L. Calvert, and M.J. Donahoo, 
IEEE/ ACM Trans. Network {\bf 5}, 770 (1997); M. Faloutsos, P. 
Faloutsos, and C. Faloutsos, Comp. Comm. Rev. {\bf 29}, 251 (1999).
\bibitem{albert} R. Albert, H. Jeong, and A.-L. Barabasi, Nature, {\bf 401}, 
130 (1999).   
\bibitem{kumar} R.P. Kumar, S. Raghavan, D. Rajalopagan, and A. S. Tomkins, 
Proceedings of the 9th ACM Symposium on Principles of Database Systems, 1.
\bibitem{redner} S. Redner, Eur. Phys. J. B {\bf 4}, 131 (1998).
\bibitem{metabolic} H. Jeong, B. Tombor, R. Albert, Z.N. Oltvani, and A.-L.
Barab\'asi, Nature {\bf 407}, 651 (2000).
\bibitem{Kleinberg} J.M. Kleinberg, R. Kumar, P. Raghavan, S. Rajagopalan, 
and A. Tomkins, Proceedings of the International Conference on Combinatorics 
and Computing. (1999). 
\bibitem{Broder} A. Broder, R. Kumar, F. Maghoul, P. Raghavan, S. Rajagopalan, 
and R. Stata, A, Tomkins and J. Wiener, Proceedings of the 9th WWW 
conference, Amsterdam, 309 (2000).
\bibitem{Mendes} S.N. Dorogovtsev and J.F.F. Mendes (cond-mat/0106144).
\bibitem{krapivsky} P.L. Krapivsky, G.J. Rodgers, and S. Redner, Phys. 
Rev. Lett. {\bf 86,} 5401 (2001). 
\bibitem{tadic} B. Tadi\'c, Physica A {\bf 293}, 273 (2001).
\bibitem{gauss} K.-I. Goh, B. Kahng, and D. Kim, (cond-mat/0108031). 
\end{thebibliography}
\end{document}